\documentclass[12pt]{iopart}
\usepackage{iopams}
\usepackage{graphicx}
\usepackage{dcolumn}
\usepackage{bm}

\begin{document}

\letter{Field-Induced Effects of Anisotropic Magnetic Interactions in SrCu$_{2}$(BO$_{3}$)$_{2}$}

\author{K Kodama$^{1}$, S Miyahara$^{2}$, M Takigawa$^{1}$ \footnote{Author to whom any correspondence
should be addressed (masashi@issp.u-tokyo.ac.jp)}, M Horvati\'{c}$^{3}$, C Berthier$^{3,4}$, F Mila$^{5}$, 
H Kageyama$^{6}$ and Y Ueda$^{1}$}

\address{$^{1}$ Institute for Solid State Physics, University of Tokyo, Kashiwanoha, Kashiwa, Chiba 277-8581, Japan}
\address{$^{2}$ Department of Physics, Aoyama Gakuin University, Sagamihara, Kanagawa 229-8558, Japan}
\address{$^{3}$ Grenoble High Magnetic Field Laboratory, CNRS and MPI-FKF, BP 166 - 38042 Grenoble, France}
\address{$^{4}$ Laboratoire de Spectrom\'{e}trie Physique, Universit\'{e} J. Fourier, BP 87 - 38402 St.-Martin d'H\`{e}res, France}
\address{$^{5}$ Institute of Theoretical Physics, Ecole Polytechnique F\'ed\'erale de Lausanne, CH-1015 Lausanne, Switzerland}
\address{$^{6}$ Department of Chemistry, Graduate School of Science, Kyoto University, Kyoto 606-8502, Japan}

\begin{abstract}
We observed a field-induced staggered magnetization in the 2D frustrated 
dimer singlet spin system SrCu$_{2}$(BO$_{3}$)$_{2}$ by $^{11}$B NMR, 
from which the magnitudes of the intradimer Dzyaloshinsky-Moriya interaction and 
the staggered $g$-tensor were determined.  These anisotropic interactions cause singlet-triplet mixing
and eliminate a quantum phase transition at the expected critical field $H_{c}$ for gap closing. 
They provide a quantitative account for some puzzling phenomena such as the onset of a uniform magnetization
below $H_{c}$ and the persistence of the excitation gap above $H_{c}$.  
The gap was accurately determined from the activation energy of the nuclear relaxation rate.    
\end{abstract}




Spin systems with singlet ground states exhibit a variety of quantum phase 
transitions in magnetic field \cite{rice021}.  A generic example is the Bose-Einstein 
condensation of triplets when the field exceeds the critical value at which 
the excitation energy vanishes.  This results in an antiferromagnetic order 
with the staggered moment perpendicular to the field, as has been observed, e.g., 
in TlCuCl$_{3}$ \cite{tanaka011}.  Another possibility is 
the formation of a superlattice of localized triplets due to repulsive interactions, 
which translates into magnetization plateaus at  fractional values of the saturated magnetization.  
The best known example is SrCu$_{2}$(BO$_{3}$)$_{2}$ with its two-dimensional network of orthogonal 
dimers of Cu$^{2+}$ ions (spin 1/2).  This material shows an excitation gap 
$\Delta_{0}$=35~K and plateaus at 1/8, 1/4, and 1/3 of the saturated magnetization 
\cite{miyahara031,kage991,onizuka001}. A magnetic superlattice at the 1/8-plateau has 
actually been observed by NMR experiments \cite{kodama022}. 
\begin{figure}[tbp]
\centering 
\includegraphics*[scale=0.6]{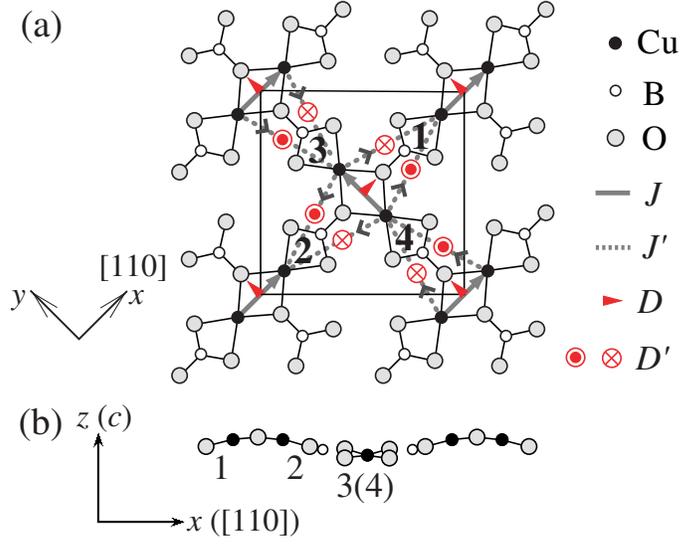}
\caption{(color online) The magnetic layer of SrCu$_{2}$(BO$_{3}$)$_{2}$ viewed
along (a) the $c$- ($z$-) and (b) the [1$\overline{1}$0]- ($y$-) directions.  Numbers
are the site indices for both Cu and B.  Symbols for $D$ and $D^\prime$ indicate the
direction of ${\bm d}$ in the Dzyaloshinsky-Moriya interaction ${\bm d} \! \cdot \!
({\bm s}_{i} \! \times \! {\bm s}_{j})$ with the bond direction $i \rightarrow j$ shown
by arrows.}
\label{crystal}
\end{figure}

These basic properties can be explained by a Heisenberg model on the 
frustrated Shastry-Sutherland lattice, 
\begin{equation}
\label{shastry}
{\cal H}_0 = J \sum _{n.n.} \bm{s}_{i} \cdot \bm{s}_{j} + J^{\prime} \sum _{n.n.n.} \bm{s}_{i} \cdot \bm{s}_{j},
\end{equation}
\noindent where $J$ and $J^{\prime}$ are the intra- and interdimer exchange interactions as shown in
\fref{crystal}a \cite{miyahara031,miyahara991}.  However, 
several aspects of low-temperature and high-field properties remain mysterious:
i) a finite magnetization appears well below the expected critical field 
for the gap closing $H_{c}=\Delta_{0}/g_{z}\mu_{B}$, where $g_{z}$ is the $g$-value along the 
field direction \cite{onizuka001, kage011}; ii) a gap seems to persist 
above $H_{c}$ \cite{nojiri991,nojiri031,tsujii031} ; iii) there appears no phase transition 
down to $T$=0 at fields below the 1/8-plateau, which means that no Bose condensation occurs, 
while one has been observed above it \cite{tsujii031};
iv) the magnetization shows a discontinuous jump at the lower boundary of the 
1/8-plateau \cite{kodama022,kage011}. The properties i) and ii) suggest that 
triplet states are mixed into the ground state by some anisotropic interactions.
The {\it interdimer} Dzyaloshinsky-Moriya (DM) interaction,
\begin{equation}
\label{interdimer}
{\cal H}^{\prime}=\sum _{n.n.n.} D^{\prime}_{ij} (s_{i}^{x}s_{j}^{y}-s_{i}^{y}s_{j}^{x}) ,
\end{equation}
where $ \mid D^{\prime}_{ij} \mid =D^{\prime}$ and the sign alternates as shown
in \fref{crystal}a, was required to explain the splitting of triplet energy levels 
\cite{cepas011,nojiri991}.  However, it does not have matrix elements between singlet and
one-triplet states.  Although the {\it intradimer} DM interaction is a candidate,  
no estimate for its magnitude has been made so far.    
  
In this Letter, we report observation of a field-induced staggered magnetization 
by $^{11}$B NMR experiments, which is also caused by singlet-triplet mixing.  
Quantitative estimates of the intradimer DM interaction and the staggered 
$g$-tensor are obtained.  Furthermore, these anisotropic interactions 
provide quantitative accounts for the above points i) and ii).  We also report  
the behavior of the gap obtained from the nuclear relaxation rate upon entering into the 1/8-plateau.  

A single crystal of SrCu$_{2}$(BO$_{3}$)$_{2}$ was grown by the
traveling-solvent-floating-zone method using LiBO$_2$ solvent \cite{kage992}.  The
NMR measurements below 18~T were performed at ISSP, University of Tokyo, while data
at higher fields were obtained using a 20MW resistive magnet at the Grenoble High
Magnetic Field Laboratory.

We first describe the anisotropic interactions compatible with the crystal structure
(space group $I\overline{4}2m$ \cite{smith911}).  A crucial feature is the
non-coplanar buckling of the magnetic CuBO$_3$ layers as depicted in
\fref{crystal}b.  This allows additional anisotropic interactions,
\begin{eqnarray}
\label{intradimer}
{\cal H}_1 & = & -\mu_B\bm{H}\cdot(\sum_{i=1}^{4}{\bf g}_{i}\cdot\bm{s}_{i}) + \nonumber  \\ 
& & D \left\{ \sum_{A} (s_{1}^{z}s_{2}^{x}-s_{1}^{x}s_{2}^{z}) -
\sum_{B} (s_{3}^{y}s_{4}^{z}-s_{3}^{z}s_{4}^{y}) \right\},
\end{eqnarray}
where A (B) denotes dimers along the $x$- ($y$) direction.
The second term, the intradimer DM interaction, is allowed since
the buckling breaks inversion symmetry at the center of dimer bonds.
The first, Zeeman term involves anisotropic $g$-tensors, which for site 1 is given
by
\[ {\bf g}_{1} = \left(
\begin{array}{ccc}
 g_{x} &  0      &  -g_{s}   \\
 0     &  g_{y}  &  0     \\
 -g_{s} &   0     &  g_{z}
\end{array}
\right) ,  \] while ${\bf g}_{2}$, ${\bf g}_{3}$, and ${\bf g}_{4}$ are obtained from ${\bf g}_{1}$ by symmetry
operations of the crystal (\fref{crystal}).  The sign of the $xz$-component
is opposite for ${\bf g}_{2}$, i.e., $g_{s}$ represents the staggered component. The
diagonal components were determined as $g_{x}$=$g_{y}$=2.05 and $g_{z}$=2.28 from
ESR measurements \cite{nojiri991}. We estimate $g_s$=0.023 by assuming
that the principal axis of ${\bf g}_{1}$ coincides with that of the electric field
gradient at Cu1 nuclei (5.6$^\circ$ from the $c$-axis \cite{kodama041}), which is
nearly perpendicular to the plane containing the four oxygen atoms surrounding one Cu1.

In magnetic fields, the anisotropic Hamiltonian ${\cal H}_1$ results in a
non-collinear magnetization consisting of four sublattices
$\bm{m}_{i}$=${\bf g}_{i}\cdot\langle\bm{s}_i\rangle\mu_{B}$ ($i$=1 to 4).  We define the
uniform and staggered moments on A and B dimers as
$\bm{m}^{A}_{u}=(\bm{m}_{1}+\bm{m}_{2})/2$,
$\bm{m}^{A}_{s}=(\bm{m}_{1}-\bm{m}_{2})/2$ and $\bm{m}^{B}_{u}=(\bm{m}_{3}+\bm{m}_{4})/2$,
$\bm{m}^{B}_{s}=(\bm{m}_{3}-\bm{m}_{4})/2$.
The crystal symmetry tells us that an external field $\bm{H}_{\rm ext} \parallel z$ 
($\bm{H}_{\rm ext} \parallel x$) induces staggered moments $\bm{m}^{A}_{s} \parallel x$ 
and $\bm{m}^{B}_{s} \parallel y$ ($\bm{m}^{A}_{s} \parallel z$). 
Such non-collinear moments, unobservable from bulk magnetization measurements, 
can be accurately detected by $^{11}$B NMR.  The magnetic
hyperfine field acting on a nuclear spin at site $i$ is expressed as
$\bm{H}^{\rm hf}_{i}$=$\sum_{j=1}^{4}{\bf A}_{ij}\cdot\bm{m}_{j}$,
where ${\bf A}_{ij}$ is the hyperfine coupling tensor between the $i$-th nuclear
spin and $\bm{m}_{j}$. The projection of $\bm{H}^{\rm hf}_{i}$ along $\bm{H}_{\rm ext}$, 
denoted as $H_{i}$, is obtained from the shift of the NMR frequency.

\begin{figure}[tbp]
\centering
\includegraphics*[scale=0.6]{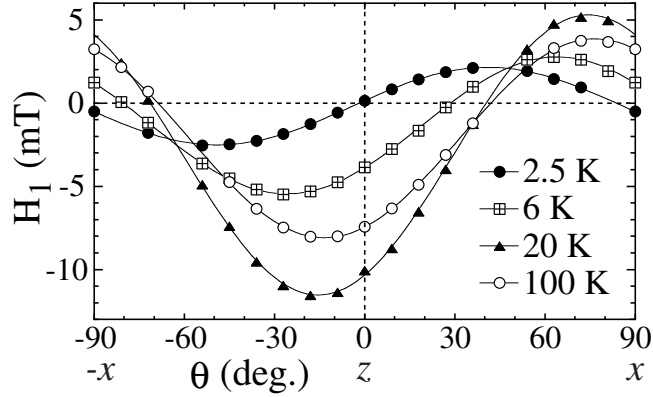}
\caption{Angular variation of $H_{1}$ at different temperatures in the external field
of 6.948~T.}
\label{angdep}
\end{figure}
\begin{figure}[tbp]
\centering
\includegraphics*[scale=0.6]{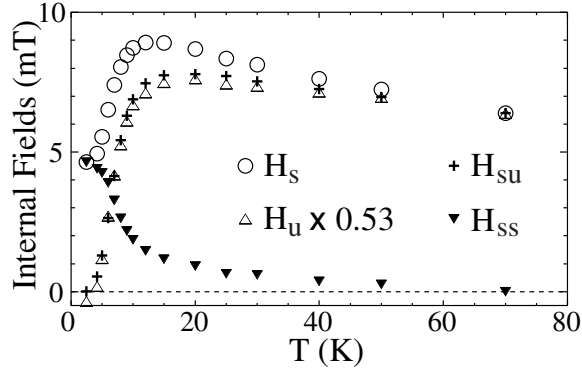}
\caption{Temperature dependences of $H_{u}$, $H_{s}$, \eref{HsHu}
$H_{su}$, and $H_{ss}$ \eref{hs} at $H_{\rm ext}$=6.948~T.}
\label{Tdep}
\end{figure}

We first discuss the results for a small field (6.948~T)  
rotated in the ($\overline{1}$10)- ($xz$-) plane.
The NMR spectra consist of several sets of quadrupole-split lines.  Each set was
assigned to a particular site based on the known anisotropy of the quadrupole
splitting \cite{kodama021}.  In \fref{angdep}, $H_1$ is plotted against the
angle $\theta$ between $\bm{H}_{\rm ext}$ and the $c$-axis at several temperatures.
By symmetry, $H_{2}(\theta)$=$H_{1}(-\theta)$ and
$H_{3}(\theta)$=$H_{4}(\theta)$ (not shown).  The $\theta$-dependence can be well
fitted by a simple sinusoidal curve (curves in \fref{angdep}),  
\begin{eqnarray}
\label{HsHu}
H_{1}+H_{2} & = & H_{u}\cos 2\theta + const \ , \nonumber  \\
H_{1}-H_{2} & = & H_{s} \sin 2\theta \ .
\end{eqnarray}
\noindent
$T$-dependences of $H_{u}$ and $H_{s}$ are plotted in \fref{Tdep}.  

The $\theta$-dependence of $H_{1}$ is not symmetric around
$\theta=0$, leading to a finite $H_{s}$.  Both $\bm{m}^{A,B}_{s}$ 
and $\bm{m}^{A,B}_{u}$ contribute to $H_{s}$
through the uniform (diagonal) and the staggered (off-diagonal) parts of
the coupling ${\bf A}_{ij}$, respectively.  If $\bm{m}^{A,B}_{u}$ and
$\bm{m}^{A,B}_{s}$ had the same $T$-dependence, this should stand for 
$H_{s}$ and $H_{u}$.  This is indeed not the case, since $H_{s}$ extrapolates
to a large finite value at $T$=0, while $H_{u}$ almost vanishes similarly to the 
magnetic susceptibility \cite{kage991}. This proves that a sizable staggered 
moment $\bm{m}^{A,B}_{s}$ is induced at $T$=0 at field values 
$H_{\rm ext} \ll H_{c}$, while the uniform moment is nearly zero.  This can be qualitatively
explained by considering ${\cal H}_{1}$ as a perturbation for an isolated dimer
\cite{miyahara041}.  Since ${\cal H}_1$ has matrix elements between singlet and
triplets on the same dimer, triplets are mixed into the ground state, resulting in
finite $\bm{m}^{A,B}_{s}$ to first order in perturbation.  In contrast,
$\bm{m}^{A,B}_{u}$ is much smaller than $\bm{m}^{A,B}_{s}$, since only higher order
terms contribute to $\bm{m}^{A,B}_{u}$.  Although field-induced staggered moments
have been reported in the Haldane chain compound NENP \cite{chiba911,fujiwara931},
this is the first observation in quasi 2D systems.    

For a quantitative analysis, we write  
\begin{eqnarray}
\label{shift}
H_{1}-H_{2} & = & 2\left( B_{xz} +  C_{xz} \right) \left( m^{A}_{uz}\sin\theta + m^{A}_{ux}\cos\theta \right)  \nonumber \\
& & + 2\left( B_{xx}-C_{xx} \right) m^{A}_{sx}\sin\theta + 2\left( B_{zz}-C_{zz} \right) m^{A}_{sz}\cos\theta  \nonumber \\
& & + 4D_{xz} \left( m^{B}_{uz}\sin\theta + m^{B}_{ux}\cos\theta \right) + 4D_{xy}m^{B}_{sy}\sin\theta ,
\end{eqnarray}
\noindent
where we have defined ${\bf B}$=${\bf A}_{11}$, ${\bf C}$=${\bf A}_{12}$, and
${\bf D}$=${\bf A}_{13}$ and used symmetry properties to obtain other
${\bf A}_{ij}$'s. The sin$2\theta$ dependence of $H_{1}$-$H_{2}$ in  
\eref{HsHu} is reproduced if the moments vary as  
$m^{A}_{sx}, m^{B}_{sy}, m^{A,B}_{uz} \propto \cos \theta$, 
and $m^{A}_{sz}, m^{A,B}_{ux} \propto \sin \theta$.
This is approximately confirmed by the numerical calculations presented below.  We define
\begin{eqnarray}
\sigma_{x} & \equiv & m^{A}_{sx}(\theta=0) = m^{B}_{sy}(\theta=0)\ , \  \sigma_{z} \equiv -m^{A}_{sz}(\theta=\pi/2) \ , \nonumber \\
M_{z} & \equiv & m^{A,B}_{uz}(\theta=0)\ , \ M^{A(B)}_{x} \equiv m^{A(B)}_{ux}(\theta=\pi/2) \ . \nonumber 
\end{eqnarray}
\noindent
Combining these relations with \eref{shift}, we obtain 
\begin{eqnarray}
\label{hs}
H_{s} & = & H_{su} + H_{ss}  \\
\label{hsu}
H_{su} & = & \left(B_{xz}\!+\!C_{xz}\!+\!2D_{xz} \right) M_{z} + 
\left(B_{xz}+ C_{xz}\right) M^{A}_{x} + 2D_{xz}M^{B}_{x}   \\
\label{hss}
H_{ss} & = & \left( B_{xx}-C_{xx}+2D_{xy} \right) \sigma_{x} - \left( B_{zz}-C_{zz} \right) \sigma_{z}. 
\end{eqnarray}

The analysis of the $^{11}$B NMR spectrum in the
1/8-plateau phase \cite{kodama042} indicates that all hyperfine couplings except
${\bf B}$ are approximately given by the classical dipolar fields that
we can calculate.  Since the diagonal components of
${\bf B}$+${\bf C}$+2${\bf D}$ were determined previously \cite{kodama021},
all the coupling parameters in the above equations are known except for $B_{xz}$.  
At high temperatures, the staggered moments $\sigma_{x}$, $\sigma_{z}$ are much smaller 
than the uniform moments $M_{z}$, $M^{A,B}_{x}$, and $M^A_x \! \approx \! M^B_x$.  
Assuming $\sigma_{x}$=$\sigma_{z}$=0 above $T$=70~K, which is also supported by 
preliminary numerical calculations, we determined $B_{xz}$ from the data of $H_{s}$ and 
the magnetization at 70~K, using \eref{hsu}.  The parameters in \eref{hsu} and 
\eref{hss} are determined as $B_{xz}$+$C_{xz}$+$2D_{xz}$=0.10, $B_{xz}$+$C_{xz}$=0.09, 
$2D_{xz}$=0.01, $B_{xx}$-$C_{xx}$+$2D_{xy}$= -0.06, and $B_{zz}$-$C_{zz}$=-0.32 (T/$\mu_{B}$).
We then obtained $H_{su}$ at other temperatures (\fref{Tdep}) from the magnetization data.  
Finally, subtracting this from $H_{s}$, $H_{ss}$ is determined in the whole temperature range 
as shown in \fref{Tdep}.  
This demonstrates that the staggered magnetization increases very steeply below 10 K and 
saturates at lower temperatures.

\begin{figure}[tbp]
\centering
\includegraphics*[scale=0.5]{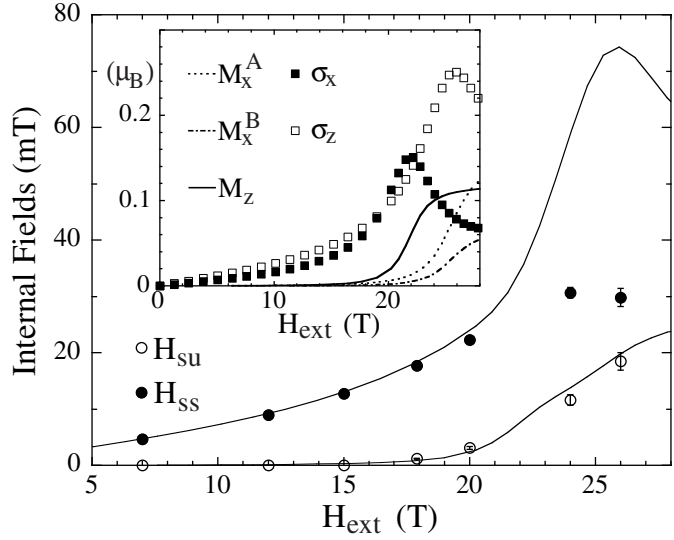}
\caption{ Field dependences of $H_{su}$ and $H_{ss}$ measured at low temperatures,
2.5~K (for $H_{\rm ext}$=7 and 12~T), 1.5~K (15 and 17.9~T), 0.64~K (20~T),
0.46~K (24~T), and 0.21~K (26~T).  The curves are obtained from the ED results
of $\sigma_{x}$, $\sigma_{z}$, $M_{z}$, and $M^{A,B}_{x}$ shown in the inset by using
\eref{hsu} and \eref{hss}.}
\label{Hdep}
\end{figure}

Measurements of $H_{i}$ were extended to higher fields up to 26~T. The magnetization data in ref.~\cite{kage011}
was used to calculate $H_{su}$ \cite{comm1}, which was then subtracted from the data of $H_{s}$ to obtain $H_{ss}$.
In the main panel of \fref{Hdep}, we plot $H_{su}$ and $H_{ss}$ in the low temperature limit.  

In order to see if these results are explained by anisotropic interactions, we 
have calculated $\bm{m}_{i}$ by exact diagonalization (ED) of the hamiltonian  
${\cal H}_0$+${\cal H}_1$+${\cal H}^{\prime}$ at $T$=0 and $H_{\rm ext}$=6.948~T 
in the ($\overline{1}$10)-plane for clusters with up to 24 sites for various values of 
$D$.  Values of other parameters are fixed: $J$=85~K, $J^{\prime}$=54~K \cite{miyahara031}, 
$D^{\prime}/J$=-0.02 \cite{cepas011},  and $g_{s}$=0.023.  By requiring that the calculated value 
of $H_{s}$ from \eref{hs} to \eref{hss} agrees with the experimental data at the lowest 
temperature (2.5~K), we obtained $D/J$=0.034.  
Taking into account the uncertainty in $g_s$ leads to
possible values of $D/J$ between 0.030 (for $g_s = 0.030$) and 0.038 (for $g_s = 0.014$).  
The choice of $D^{\prime}$ has little effect on the calculated values of $\bm{m}_{i}$.

Using the same parameter values, we have also calculated $\bm{m}_{i}$ at higher fields, as shown in the inset of 
in \fref{Hdep}.  The agreement between the calculated results of $H_{su}$ and $H_{ss}$ shown by the lines 
in the main panel of \fref{Hdep} and the experimental data is satisfactory up to about 20~T.  
Thus we now have a {\it quantitative} explanation for the development of a uniform magnetization 
at fields as low as 18~T.  Note that the staggered moment is sizable even at low fields 
($\sim 0.03~\mu_{B}/$Cu at 15~T, \fref{Hdep} inset), where the uniform moment is negligible.  
The experimental and theoretical results of $H_{ss}$ deviate at higher fields, 
probably due to finite size effects as discussed below. 

\begin{figure}[tbp]
\centering
\includegraphics*[scale=0.5]{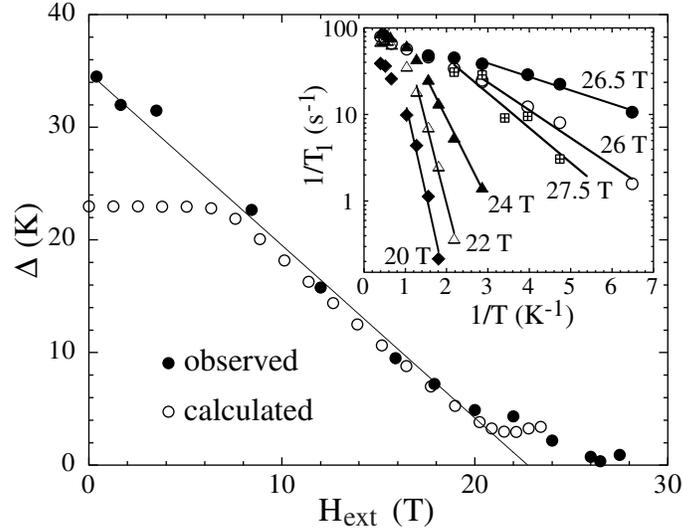}
\caption{Field dependence of the excitation gap (solid circles in the main panel)
determined from the activated behavior of 1/$T_1$ (inset) are compared with
the results of ED calculation (open circles).  The data at 27.5~T, where NMR
spectrum show many lines in the 1/8-plateau phase,
were obtained for the line with the largest (negative) hyperfine field.}
\label{t1gap}
\end{figure}
Let us now discuss the dynamics.  The nuclear spin-lattice relaxation rate (1/$T_1$)
was measured for $\bm{H}_{\rm ext} \parallel \bm{c}$ up to 27.5~T.  It shows an
activated behavior at low temperatures (\fref{t1gap} inset).  The activation
gap ($\Delta$) shown in \fref{t1gap} follows a linear $H$-dependence below 20~T,
$\Delta=\Delta_{0}-g_{z}\mu_{B}H$ with $\Delta_{0}$=34.8~K and $g_{z}$=2.28
as expected.  At higher fields, however, it deviates from this relation
and remains finite even above the expected critical field
$H_{c}$=$\Delta_{0}/g_{z}\mu_{B}$=22.7~T.  The gap reaches a minimum at
26.5~T, which is at the boundary to the 1/8-plateau, and increases
again inside the 1/8-plateau (27.5~T).  The finite gap above $H_{c}$,
which naturally explains the absence of Bose condensation,  
is consistent with earlier reports.  For example, the gap was estimated 
from the specific heat data \cite{tsujii031} as $\Delta$=4.6 (3.2)~K at 
$H_{\rm ext}$=22 (24)~T.  The ESR data \cite{nojiri031} also
shows a similar deviation from the linear $H$-dependence
of the lowest triplet energy near $H_{c}$.

One expects that the singlet-triplet mixing due to ${\cal H}_{1}$ will prevent the gap 
from closing, as has been discussed for NENP \cite{mitra941}.
We have confirmed this by numerical calculations.  The energy of
the lowest excited state obtained by ED calculations using the same parameters
is plotted in \fref{t1gap}.  The constant value 
at low fields is due to the singlet bound state of two triplets having lower energy
than one triplet \cite{knetter001}.  However, they do not contribute to the nuclear relaxation.
Near 22~T, the results clearly indicate level repulsion due to mixing,
reproducing the experimental behavior. Let us note one subtle point.
The interdimer DM interaction ${\cal H}^{\prime}$ splits the one-triplet
excitations into two branches \cite{cepas011}.  Only one of them is mixed with
the singlet by ${\cal H}_{1}$ and this branch must have a lower energy
for the level repulsion to occur.  This determines the sign of $D^{\prime}$ to be negative.  
Note that the sign of $D^{\prime}$ was not known before.

Finally, the theory is not applicable when the field is
too close to the 1/8-plateau.  The cluster size of our calculation is not large enough 
to take proper account of interaction between triplets, which should become important 
near the 1/8-plateau.  

In conclusion, we have shown that a large staggered magnetization is induced 
by the magnetic field in the presence of the intradimer DM interaction and 
staggered $g$-tensor, whose magnitudes are obtained as $0.03 \leq D/J \leq 0.038$ for 
$0.014 \leq g_s \leq 0.03$.  The singlet-triplet mixing and level repulsion 
caused by these interactions account for the finite uniform magnetization below $H_{c}$ and 
the persistence of the excitation gap above $H_{c}$.  They thus turn the quantum phase transition  
at $H_{c}$ into a crossover. This is the first example 
of such field-induced phenomena in quasi 2D spin systems.  
Our determination of the magnitudes and sign of the 
DM interactions should set the stage for further investigation of other remarkable properties of 
SrCu$_{2}$(BO$_{3}$)$_{2}$.  In particular, the signature of a phase transition in the specific heat 
observed above the 1/8-plateau by Tsujii {\it et al.} \cite{tsujii031} calls for a precise understanding 
of the interplay between the DM interaction and other bosonic interaction such as 
bound state formation \cite{knetter001} and correlated hopping \cite{momoi001}.   

\ack
We thank M.~Oshikawa, K.~Ueda, and T.~Ziman for stimulating discussions.
This work is supported by the Grants--in--aid for Scientific Research 
from the MEXT Japan and by the Swiss National Fund.


\begin{thebibliography}{99}
\bibitem{comm1} We estimated $M^{A}_{x}$ and $M^{B}_{x}$ separately by using 
the magnetization data of $(M^{A}_{x}+M^{B}_{x})/2$ and assuming the ratio
$M^{A}_{x}/M^{B}_{x}$=2.7 for the field range 18 - 26~T as indicated by the 
ED calculations (\fref{Hdep} inset).
\end{thebibliography}
\end{document}